\documentclass[aps,prl,twocolumn,superscriptaddress]{revtex4}%
\usepackage{epsfig,dsfont,amssymb,amsmath,amsthm,amsfonts,amsbsy,mathrsfs}
\usepackage{graphicx}
\usepackage{amsmath}
\usepackage{amssymb}%
\begin{document}
\title{First-Principles Study of Hybrid Graphene and MoS$_2$ Nanocomposites}

\author{Wei Hu}
\thanks{Corresponding author. E-mail: whu@lbl.gov}
\affiliation{Computational Research Division, Lawrence Berkeley
National Laboratory, Berkeley, CA 94720, USA} \affiliation{Hefei
National Laboratory for Physical Sciences at Microscale, and
Synergetic Innovation Center of Quantum Information and Quantum
Physics, University of Science and Technology of China, Hefei, Anhui
230026, China}

\author{Tian Wang}
\affiliation{Department of Precision Machinery and Precision
Instrumentation, University of Science and Technology of China,
Hefei, Anhui 230026, China}

\author{Jinlong Yang}
\thanks{Corresponding author. E-mail: jlyang@ustc.edu.cn}
\affiliation{Hefei National Laboratory for Physical Sciences at
Microscale, and Synergetic Innovation Center of Quantum Information
and Quantum Physics, University of Science and Technology of China,
Hefei, Anhui 230026, China}

\date{\today}

\pacs{ }

\begin{abstract}

Combining the electronic properties of graphene and molybdenum
disulphide (MoS$_2$) monolayers in two-dimensional (2D) ultrathin
hybrid nanocomposites have been synthesized experimentally to create
excellent electronic, electrochemical, photovoltaic, photoresponsive
and memory devices. Here, first-principles calculations are
performed to investigate the electronic, electrical and optical
properties in hybrid G/MoS$_2$ and G/MoS$_2$/G nanocomposites. It
turns out that weak van der Waals interactions dominate between
graphene and MoS$_2$ with their intrinsic electronic properties
preserved. Interestingly, tunable p-type doping of graphene is very
easy to achieve by applying electric fields perpendicular to hybrid
G/MoS$_2$ and G/MoS$_2$/G nanocomposites, because electrons can
easily transfer from the Dirac point of graphene to the conduction
band of MoS$_2$ due to the work function of graphene close to the
electronic affinity of MoS$_2$. Vertical electric fields can
generate strong p-type but weak n-type doping of graphene, inducing
electron-hole pairs in hybrid G/MoS$_2$/G sandwiched nanocomposites.
Moreover, improved optical properties in hybrid G/MoS$_2$ and
G/MoS$_2$/G nanocomposites are also expected with potential
photovoltaic and photoresponsive applications.

\end{abstract}

\maketitle

\section{Introduction}

Graphene, a two-dimensional (2D) sp$^2$-hybridized carbon sheet, has
received considerable interest recently owing to its outstanding
properties,\cite{Scinece_306_666_2004, NatureMater_6_183_2007,
RMP_81_109_2009, NatPhotonics_4_611_2010} especially, high carrier
mobility, with great potential applications for graphene-based
electronic devices, such as field effect transistors (FETs). But,
intrinsic electronic properties of graphene depend sensitively on
the substrates due to strong graphene-substrate interactions, such
as SiO$_2$,\cite{NanoLett_7_1643_2007, NatPhys_4_144_2008,
PRL_106_106801_2011} SiC,\cite{NatureMater_6_770_2007,
PRL_99_126805_2007, PRL_108_246104_2012} and metal
surfaces.\cite{PRL_101_026803_2008, PRB_79_195425_2009,
NatureNanotech_6_179_2011} Finding an ideal substrate for graphene
remains a significant challenge.

Interestingly, many 2D ultrathin hybrid graphene-based
nanocomposites have been widely studied experimentally and
theoretically, such as graphene/hexagonal born nitride
(G/h-BN),\cite{Nature_5_722_2010, NatureMater_10_282_2011,
NanoLett_12_714_2012} graphene/graphitic carbon nitride
(G/g-C$_3$N$_4$),\cite{JPCC_115_7355_2011, JACS_133_8074_2011,
JACS_134_4393_2012} and graphene/graphitic ZnO
(G/g-ZnO).\cite{JCP_138_124706_2013, JPCC_117_10536_2013,
JAP_113_054307_2013} These hybrid graphene-based nanocomposites show
much more new properties far beyond their simplex components.
Furthermore, most of them are ideal substrates for graphene to
preserve its intrinsic electronic properties.

Recently, 2D flexible heterostructures consisting of graphene and
monolayer molybdenum disulphide\cite{PRL_87_196803_2001,
PRL_105_136805_2010, NanoLett_10_1271_2010,
NatureNanotechnol_6_147_2011, ACSNano_6_74_2012} (G/MoS$_2$) have
been synthesized experimentally\cite{CC_47_4252_2011,
ACSNano_5_4720_2011, Science_335_947_2012, NanoLett_12_2784_2012,
JACS_134_6575_2012, NatureMater_12_246_2013,
NatureCommun_4_1624_2013, ACSNano_7_3246_2013,
NatureNanotechnol_8_826_2013, Nanoscale_6_1071_2014} with great
applications in excellent electronic, electrochemical, photovoltaic,
photoresponsive and memory devices. Monolayer MoS$_2$ itself is an
interesting semiconducting transition metal dichalcogenide, which
has been widely studied experimentally and
theoretically\cite{PRL_87_196803_2001, PRL_105_136805_2010,
NanoLett_10_1271_2010, NatureNanotechnol_6_147_2011,
ACSNano_6_74_2012} due to its outstanding structural, electronic and
optical properties superior to other 2D materials as graphene
substrates. For example, monolayer MoS$_2$ has a appropriate bandgap
(1.90 $eV$)\cite{PRL_105_136805_2010} to achieve a high current
on/off ratio of 10000 in G/MoS$_2$ based FETs far superior to
vertical G/h-BN heterostructures with an insufficient current on/off
ratio of 50 due to the large bandgap (5.97 $eV$) of
h-BN.\cite{Science_335_947_2012} Most recently, hybrid G/MoS$_2$/G
sandwiched nanocomposites have also been synthesized
experimentally,\cite{Science_340_1311_2013, ACSNano_7_7021_2013}
showing strong light-matter interactions and large quantum tunneling
current modulations in flexible graphene-based FETs superior to
hybrid G/MoS$_2$ nanocomposites. Theoretically, only few
works\cite{Nanoscale_3_3883_2011, JPCC_117_15347_2013} have been
focused on the structural and electronic of hybrid G/MoS$_2$
nanocomposites. Therefore, a systematic theoretical work on hybrid
graphene and MoS$_2$ nanocomposites (G/MoS$_2$ and G/MoS$_2$/G) is
very desirable with more exciting new properties to be expected,
such as electrical and optical properties.

In the present work, we study structural, electronic, electrical and
optical properties in hybrid G/MoS$_2$ and G/MoS$_2$/G
nanocomposites using first-principles calculations. Graphene
interacts overall weakly with  via van der Waals interactions with
their intrinsic electronic properties preserved. Applying vertical
electric fields is very easy to induce tunable p-type doping of
graphene in hybrid G/MoS$_2$ nanocomposites and generate strong
p-type but weak n-type doping of graphene in hybrid G/MoS$_2$/G
sandwiched nanocomposites. Moreover, hybrid graphene and MoS$_2$
nanocomposites show enhanced optical absorption compared to simplex
graphene and MoS$_2$ monolayers.

\section{Theoretical Models and Methods}

The lattice parameters of graphene and MoS$_2$ monolayers calculated
to setup unit cell are $a$(G) = 2.47 {\AA} and $a$(MoS$_2$) = 3.19
{\AA}.\cite{Nanoscale_3_3883_2011, JPCC_117_15347_2013} In order to
simulate hybrid graphene and MoS$_2$ nanocomposites, a 2$\sqrt{3}$
$\times$ 2$\sqrt{3}$ supercell of graphene (24 carbom atoms) is used
to match a $\sqrt{7}$ $\times$ $\sqrt{7}$ supercell of MoS$_2$ (7
sulfur and 14 molybdenum atoms) with a smaller lattice mismatch of
about 1\% compared with previous theoretical
studies.\cite{Nanoscale_3_3883_2011, JPCC_117_15347_2013}
Furthermore, our designed hybrid graphene and MoS$_2$ nanocomposites
are consistent with recent experimental observation of a faint
moir\'{e} structure with high resolution STM images of graphene on
MoS$_2$.\cite{Nanoscale_6_1071_2014} In this work, both hybrid
G/MoS$_2$ and G/MoS$_2$/G nanocomposites are considered as shown in
FIG 1. The vacuum space in the $Z$ direction is about 15 {\AA} to
separate the interactions between neighboring slabs.

\begin{figure}[htbp]
\begin{center}
\includegraphics[width=0.5\textwidth]{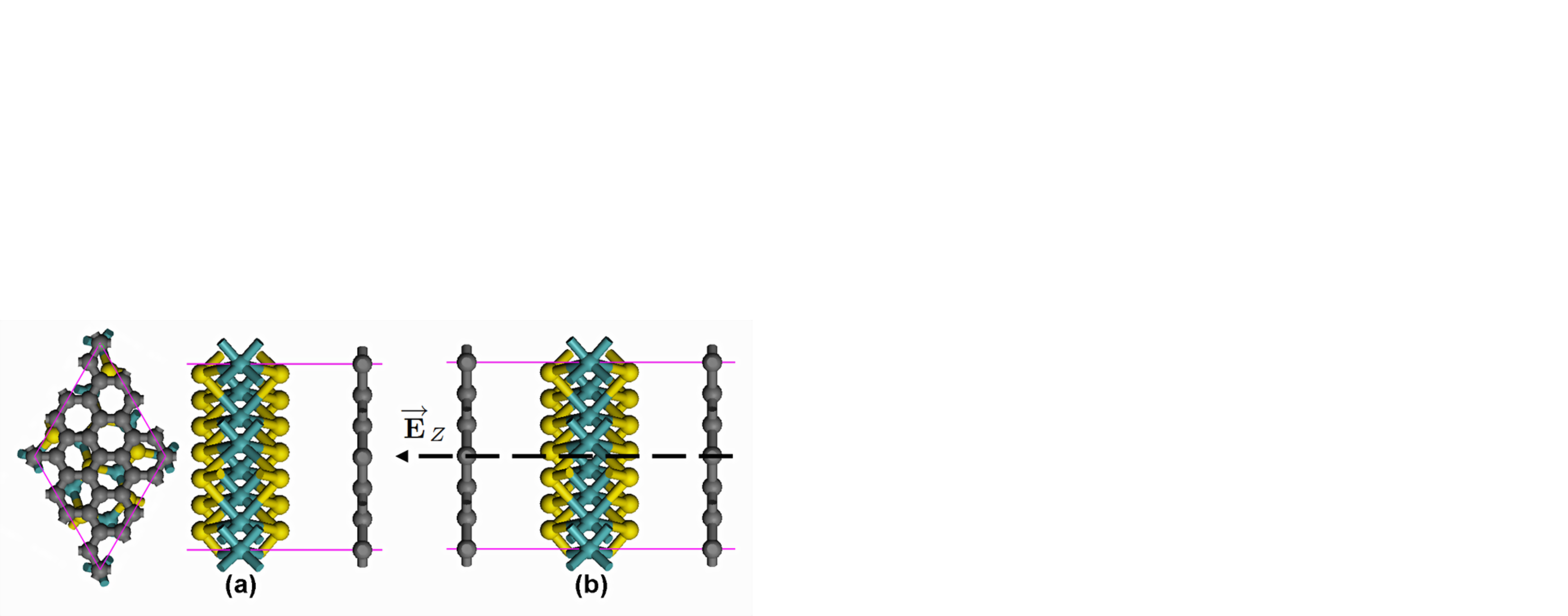}
\end{center}
\caption{(Color online) Atomic geometries of hybrid (a) G/MoS$_2$
and (b) G/MoS$_2$/G nanocomposites. A vertical electric field is
applied perpendicular to the layers. The gray, yellow and blue balls
denote carbon, sulfur and molybdenum atoms, respectively. }
\end{figure}

First-principles calculations are based on the density functional
theory (DFT) implemented in the VASP
package.\cite{PRB_47_558_1993_VASP} The generalized gradient
approximation of Perdew, Burke, and Ernzerhof
(GGA-PBE)\cite{PRL_77_1996} with van der Waals (vdW) correction
proposed by Grimme (DFT-D2)\cite{JCC_27_1787_2006_Grimme} is chosen
due to its good description of long-range vdW
interactions.\cite{JPCC_111_11199_2007, PCCP_10_2722_2008,
NanoLett_11_5274_2011, PRB_83_245429_2011, PRB_85_125415_2012,
PRB_85_235448_2012, PCCP_14_8179_2012, PCCP_15_497_2013,
PCCP_15_5753_2013, Nanoscale_5_9062_2013, JPCL_4_2158_2013,
PCCP_2014} As an benchmark, DFT-D2 calculations give a good bilayer
distance of 3.25 {\AA} and binding energy of -25 $meV$ per carbon
atom for bilayer graphene, fully agreeing with previous
experimental\cite{PR_100_544_1955, PRB_69_155406_2004} and
theoretical\cite{PRB_85_205402_2012, JCP_138_054701_2013} studies.
The energy cutoff is set to be 500 $eV$. The surface Brillouin zone
is sampled with a 5 $\times$ 5 regular mesh and 240 $k$ points are
used for calculating the tiny band gaps at the Dirac points of
silicene. All the geometry structures are fully relaxed until energy
and forces are converged to 10$^{-5}$ $eV$ and 0.01 $eV$/{\AA},
respectively. Dipole correction is employed to cancel the errors of
electrostatic potential, atomic forces and total energy, caused by
periodic boundary condition.\cite{PRB_51_4014_1995_DipoleCorrection}
The external electric field is introduced in the VASP by the dipole
layer method with the dipole placed in the vacuum region of periodic
supercell.

To study the optical properties of hybrid graphene and MoS$_2$
nanocomposites, the frequency-dependent dielectric matrix is
calculated.\cite{PRB_73_045112_2006} The imaginary part of
dielectric matrix is determined by a summation over states as
\begin{equation}
\begin{aligned}
\varepsilon_{\alpha\beta}^{\prime\prime}=&\frac{4\pi^2e^2}{\Omega}\mathop{\lim}\limits_{q \to 0}\frac{1}{q^2}\sum_{c,v,\textbf{k}}2w_{\textbf{k}}\delta(\epsilon_{c\textbf{k}}-\epsilon_{v\textbf{k}}-\omega) \\
                               &\times\langle\mu_{c\textbf{k}+\textbf{e}_{_{\alpha}}q}|\mu_{v\textbf{k}}\rangle\langle\mu_{c\textbf{k}+\textbf{e}_{_{\beta}}q}|\mu_{v\textbf{k}}\rangle^{*}
\end{aligned}
\end{equation}
where, $\Omega$ is the volume of the primitive cell,
$w_{\textbf{k}}$ is the $\textbf{k}$ point weight, $c$ and $v$ are
the conduction and valence band states respectively,
$\epsilon_{c\textbf{k}}$ and $\mu_{c\textbf{k}}$ are the eigenvalues
and wavefunctions at the $\textbf{k}$ point respectively, and
$\textbf{e}_{_{\alpha}}$ are the unit vectors for the three
Cartesian directions. In order to accurately calculate the optical
properties of hybrid graphene and MoS$_2$ nanocomposites, a large 21
$\times$ 21 regular mesh for the surface Brillouin zone, a large
number of empty conduction band states (three times more than the
number of valence band) and frequency grid points (4000) are
adopted. Note that the optical properties of pristine graphene and
MoS$_2$ monolayers are crosschecked and consistent with previous
theoretical calculations.\cite{JCP_139_154704_2013,
PRB_88_045412_2013}

In order to evaluate the stability of hybrid graphene and MoS$_2$
nanocomposites, the interface binding energy is defined as
\begin{equation}
E_{b}=E(G/MoS_2)-E(G)-E(MoS_2)
\end{equation}
where, $E$$(G/MoS_2)$, $E$$(G)$ and $E$$(MoS_2)$ represent the total
energy of hybrid graphene and MoS$_2$ nanocomposites, pristine
graphene and MoS$_2$ monolayers, respectively.

\section{Results and Discussion}

First, we check the structural and electronic properties of pristine
graphene and MoS$_2$ monolayers, agreeing well with previous
theoretical studies.\cite{Nanoscale_3_3883_2011,
JPCC_117_15347_2013} Pristine graphene monolayer is a zero-gap
semiconductor, showing a linear Dirac-like dispersion relation
$E$($k$) = $\pm$$\hbar$$\nu_{F}$$|$$k$$|$ around the Fermi level,
where $\nu$$_{F}$ is the Fermi velocity, and $\nu$$_{F}$$(G)$ =
0.8$\times$10$^6$ $m/s$\cite{JCP_139_154704_2013} at the Dirac point
of graphene, although GGA-PBE calculations underestimate the Fermi
velocity of graphene by 15$\sim$20\%.\cite{PRB_76_205411_2007}
Pristine MoS$_2$ monolayer is a semiconductor with a direct band gap
of 1.64 $eV$, although GGA-PBE calculations\cite{PRB_78_235104_2008}
slightly underestimate this band gap value (1.90
$eV$).\cite{PRL_105_136805_2010}

We then study the structural and electronic properties of hybrid
graphene and MoS$_2$ nanocomposites as summarized in TABLE I. The
equilibrium spacings of 3.37 and 3.36 {\AA} are obtained for hybrid
G/MoS$_2$ and G/MoS$_2$/G nanocomposites with corresponding binding
energy of -20.5 and 27.0 $meV$ per atom, respectively. Thus,
graphene is physically adsorbed on monolayer MoS$_2$ via weak van
der Waals (vdW) interactions, and intrinsic electronic properties of
graphene and MoS$_2$ can be preserved in ultrathin hybrid
nanocomposites, agreeing well with previous
experimental\cite{CC_47_4252_2011, ACSNano_5_4720_2011,
Science_335_947_2012, NanoLett_12_2784_2012, JACS_134_6575_2012,
NatureMater_12_246_2013, NatureCommun_4_1624_2013,
ACSNano_7_3246_2013, NatureNanotechnol_8_826_2013,
Nanoscale_6_1071_2014, Science_340_1311_2013, ACSNano_7_7021_2013}
and theoretical\cite{Nanoscale_3_3883_2011, JPCC_117_15347_2013}
studies.

\begin{table}
\caption{DFT-D2 calculated equilibrium interfacial distance $D_0$
({\AA}) and binding energy per carbon atom $E_b$ ($meV$) in hybrid
G/MoS$_2$ and G/MoS$_2$/G nanocomposites.}
\begin{tabular}{ccccc} \\ \hline \hline
DFT-D2 &  $D_0$  &  $E_b$  \ \\
\hline
G/MoS$_2$   &  3.37  &  -20.5  \ \\
G/MoS$_2$/G   &  3.36/3.36  &  -27.0  \ \\
\hline \hline
\end{tabular}
\end{table}

Electronic band structures of hybrid graphene and MoS$_2$
nanocomposites are shown as FIG 2. We find that linear Dirac-like
dispersion relation around the Fermi levels of graphene is still
preserved in hybrid G/MoS$_2$ and G/MoS$_2$/G nanocomposites, though
tiny band gaps (1 $meV$) are opened at the Dirac points of graphene,
which are significantly lower than thermal fluctuation (about 25
$meV$) at room temperature and trend to vanish in experiments. Note
that induced graphene band gaps are typically sensitive to other
external conditions, such as interlayer
separation,\cite{JCP_139_154704_2013} showing that the band gap
values increase gradually with the interlayer separation decrease,
thus tunable, with a potential for flexible graphene-based FETs.

\begin{figure}[htbp]
\begin{center}
\includegraphics[width=0.5\textwidth]{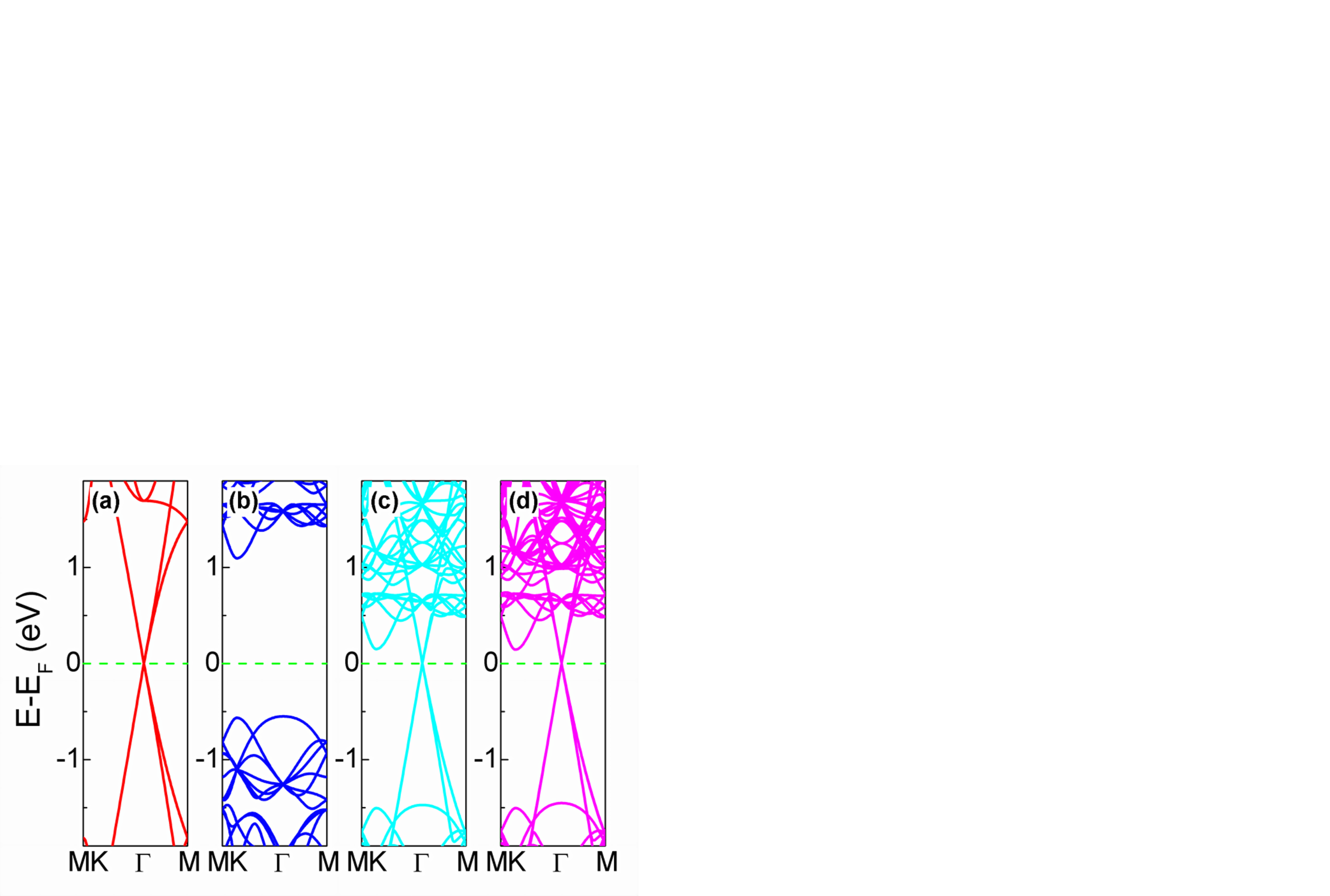}
\end{center}
\caption{(Color online) Electronic band structures of (a) graphene
and (b) MoS$_2$ as well as corresponding hybrid (c) G/MoS$_2$ and
(d) G/MoS$_2$/G nanocomposites. The Fermi level is set to zero and
marked by green dotted lines.}
\end{figure}

High-performance field-effect tunneling transistors have been
achieved experimentally\cite{Science_335_947_2012,
NatureCommun_4_1624_2013, NatureNanotechnol_8_826_2013,
Science_340_1311_2013, ACSNano_7_7021_2013} in hybrid graphene and
MoS$_2$ nanocomposites. Thus, the electronic properties of hybrid
G/MoS$_2$ and G/MoS$_2$/G nanocomposites affected by applying
vertical electric fields are very desirable as shown in FIG 3.
Interestingly, negative vertical electric fields can induce p-type
doping of graphene in hybrid G/MoS$_2$ nanocomposites. But, positive
electric fields almost have on effect on the electronic properties
of hybrid G/MoS$_2$ nanocomposites. This is because electrons can
easily from the Dirac point of graphene to the conduction band of
MoS$_2$ but difficulty from the valence band of MoS$_2$ to the Dirac
point of graphene due to the work function (4.3
$eV$)\cite{JCP_138_124706_2013} of graphene close to the electronic
affinity (4.2 $eV$)\cite{NatureCommun_4_1624_2013} of monolayer
MoS$_2$. Interestingly, vertical electric fields can generate strong
p-type but weak n-type doping of graphene at both negative and
positive electric fields due to the symmetry in hybrid G/MoS$_2$/G
nanocomposites. Based on the linear dispersion around the Dirac
point of graphene,\cite{RMP_81_109_2009} the charge carrier (hole or
electron) concentration of doped graphene can be estimated by the
following equation\cite{JCP_138_124706_2013}
\begin{equation}
N_{h/e}=\frac{(\bigtriangleup{E_D})^2}{\pi(\hbar\nu_{F})^{2}}
\end{equation}
where $\bigtriangleup$$E_D$ is the shift of graphene's Dirac point
($E_D$) relative to the Fermi level ($E_F$), that is
$\bigtriangleup$$E_D$ = $E_D$ - $E_F$. The calculated charge carrier
concentrations in hybrid G/MoS$_2$ and G/MoS$_2$/G nanocomposites
are shown in FIG 4. These values are more than 3 orders of magnitude
larger than the intrinsic charge carrier concentration of graphene
at room temperature ($n$ = $\pi$$k_{B}^2$$T^2$/6$\hbar$$\nu_{F}^2$ =
6$\times$$10^{10}$ $cm^{-2}$).\cite{PRL_108_246104_2012}
Furthermore, the doping charge carrier concentrations of graphene in
hybrid nanocomposites are increased with the vertical electric
fields. Therefore, the field-effect in hybrid graphene and MoS$_2$
nanocomposites is effective and tunable for high-performance FETs
and p-n junctions.\cite{Science_340_1311_2013, ACSNano_7_7021_2013}

\begin{figure}[htbp]
\begin{center}
\includegraphics[width=0.5\textwidth]{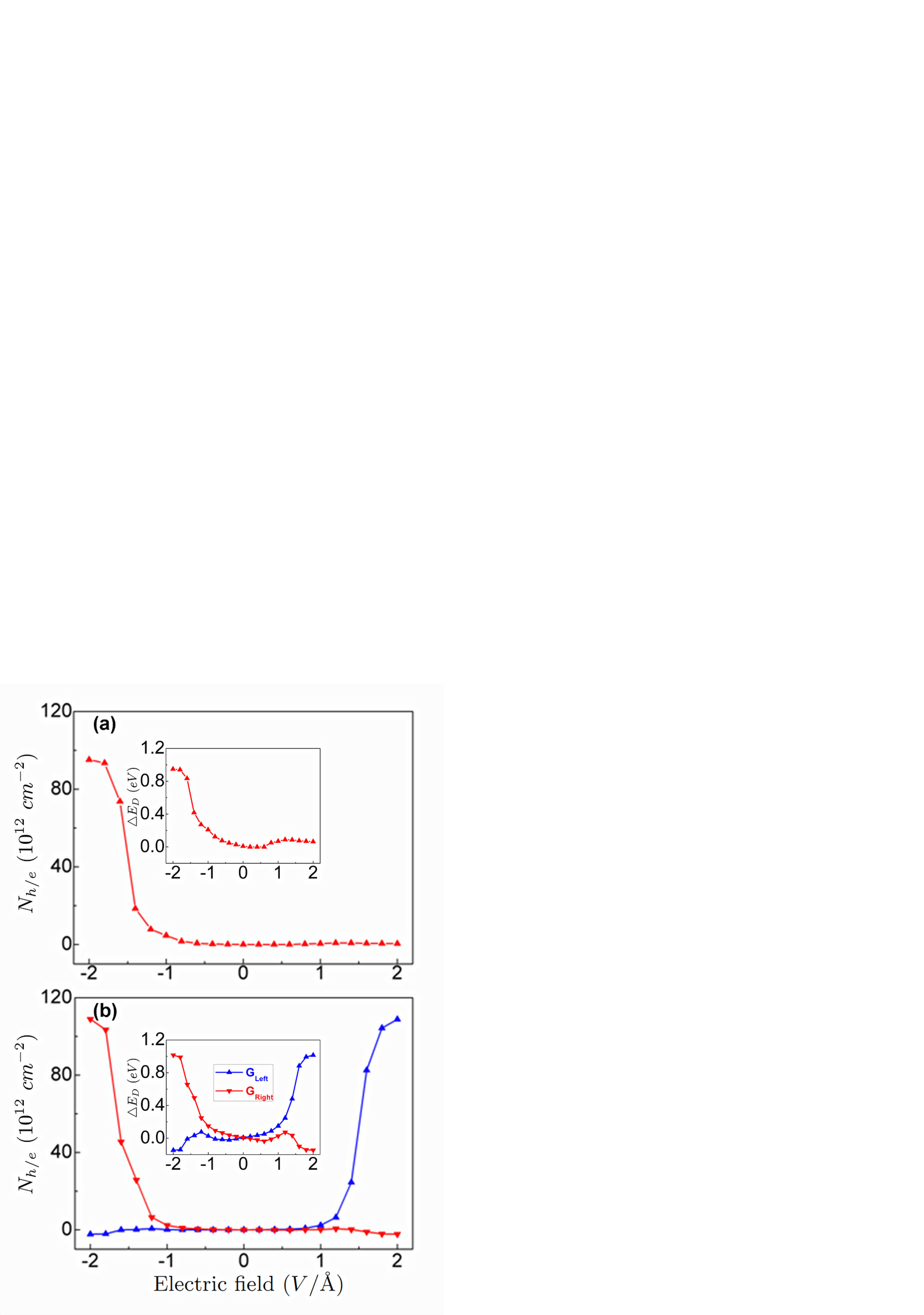}
\end{center}
\caption{(Color online) The doping charge carrier concentrations
$N_{h/e}$ ($10^{12}$ $cm^{-2}$) of graphene in hybrid G/MoS$_2$ and
G/MoS$_2$/G nanocomposites as a function of vertical electric field
$E$ ($V$/{\AA}). The energy shift $\bigtriangleup$$E_D$ ($eV$) of
graphene's Dirac point relative to the Fermi level is shown in the
inset.}
\end{figure}

As an example, electronic band structures of hybrid G/MoS$_2$ and
G/MoS$_2$/G nanocomposites and their corresponding XY-averaged
electrostatic potential ($\textbf{V}$) in the $Z$ direction at a
vertical electric field of -2.0 $V$/{\AA} are shown in FIG 4.
Charge-transfer complexes are formed in hybrid graphene and MoS$_2$
nanocomposites affected by vertical electric fields. The
differential charge density ($\triangle$$\rho$ = $\rho$(G/MoS$_2$) -
$\rho$(G) - $\rho$(MoS$_2$)) of hybrid G/MoS$_2$ and G/MoS$_2$/G
nanocomposites are shown in FIG 3c. Particularly, electron-hole
pairs are well separated in hybrid G/MoS$_2$/G sandwiched
nanocomposites with more excellent
applications\cite{Science_340_1311_2013, ACSNano_7_7021_2013}
compared with hybrid G/MoS$_2$ nanocomposites. Note that the band
gap values at graphene's Dirac points in hybrid G/MoS$_2$ and
G/MoS$_2$/G nanocomposites are almost not affected by vertical
electric fields, different to bilayer
graphene.\cite{Nature_459_820_2009, NanoLett_10_426_2010}

\begin{figure}[htbp]
\begin{center}
\includegraphics[width=0.5\textwidth]{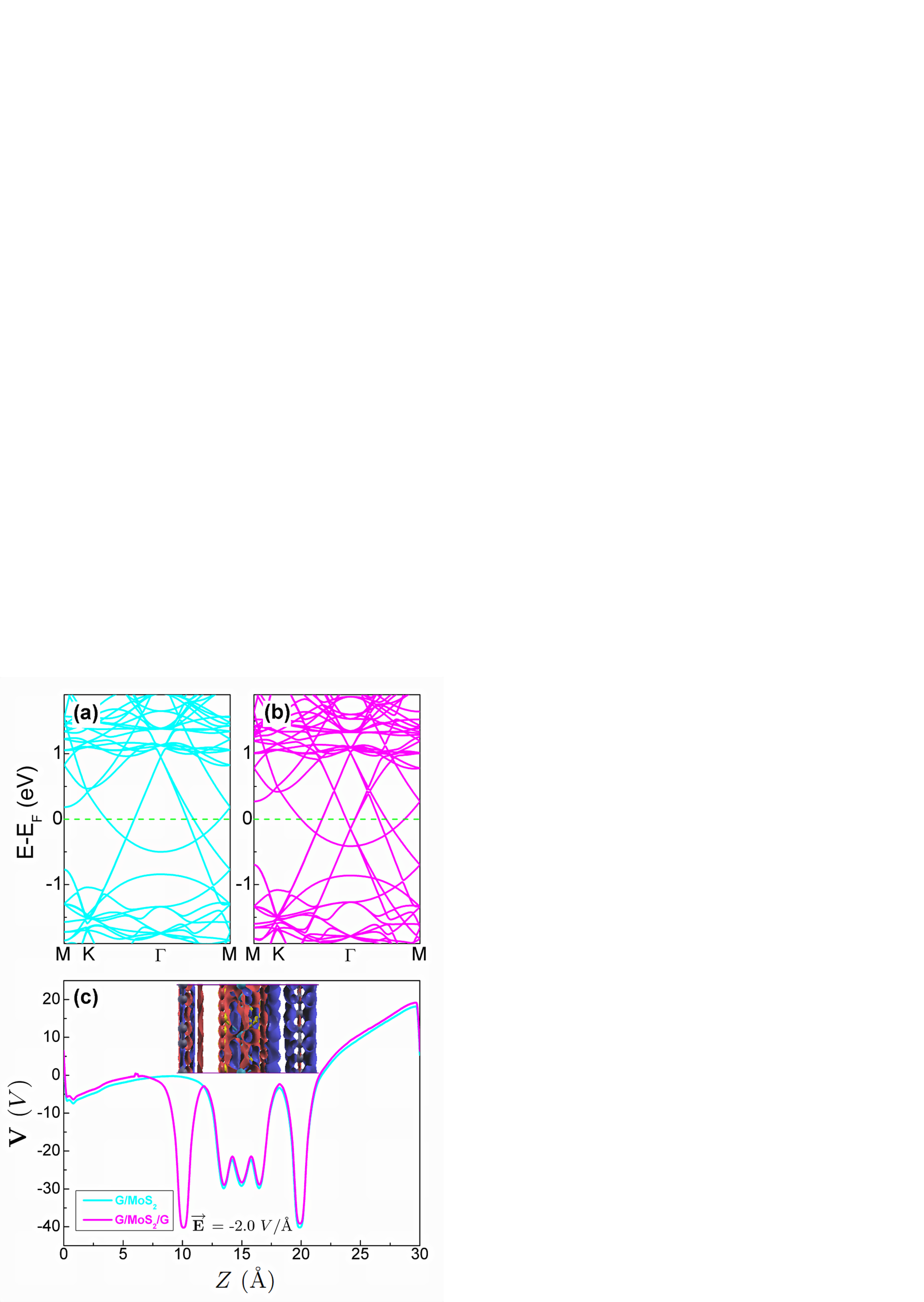}
\end{center}
\caption{(Color online) Electronic band structures of hybrid (a)
G/MoS$_2$ and (b) G/MoS$_2$/G nanocomposites and (c) their
corresponding XY-averaged electrostatic potential $\textbf{V}$ ($V$)
in the $Z$ ({\AA}) direction at a vertical electric field of -2.0
$V$/{\AA}. Differential charge density with a isosurface value of
0.001 $e$/{\AA}$^3$ of hybrid G/MoS$_2$/G nanocomposites is shown in
the insert. The red and blue regions indicate electron increase and
decrease, respectively.}
\end{figure}

Besides commonly focused electronic structures in hybrid G/MoS$_2$
nanocomposites,\cite{Nanoscale_3_3883_2011, JPCC_117_15347_2013} we
also study the optical properties in hybrid G/MoS$_2$ and
G/MoS$_2$/G nanocomposites due to their photovoltaic and
photoresponsive applications.\cite{ACSNano_7_3246_2013,
NatureNanotechnol_8_826_2013, Nanoscale_6_1071_2014,
Science_340_1311_2013} Note that pristine graphene and MoS$_2$
monolayers themselves display outstanding optical
properties,\cite{NatPhotonics_4_611_2010, NanoLett_10_1271_2010} but
interlayer interaction and charge transfer in graphene-based hybrid
nanocomposites can induce new optical
transitions.\cite{JACS_134_4393_2012} In optical property
calculations, the imaginary part of dielectric function for pristine
graphene and MoS$_2$ monolayers as well as corresponding hybrid
graphene and MoS$_2$ nanocomposites are evaluated as shown in FIG 5.
In fact, monolayer MoS$_2$ shows much stronger optical absorption
than graphene in the visible light region (200 $\sim$ 800 $nm$).
Moreover, hybrid graphene and MoS$_2$ nanocomposites exhibit more
strongly enhanced light response, especially hybrid G/MoS$_2$/G
sandwiched heterostructures, compared with simplex graphene and
MoS$_2$ monolayers, because electrons can now be easily and directly
excited from the Dirac point of graphene to the conduction band of
MoS$_2$.

\begin{figure}[htbp]
\begin{center}
\includegraphics[width=0.5\textwidth]{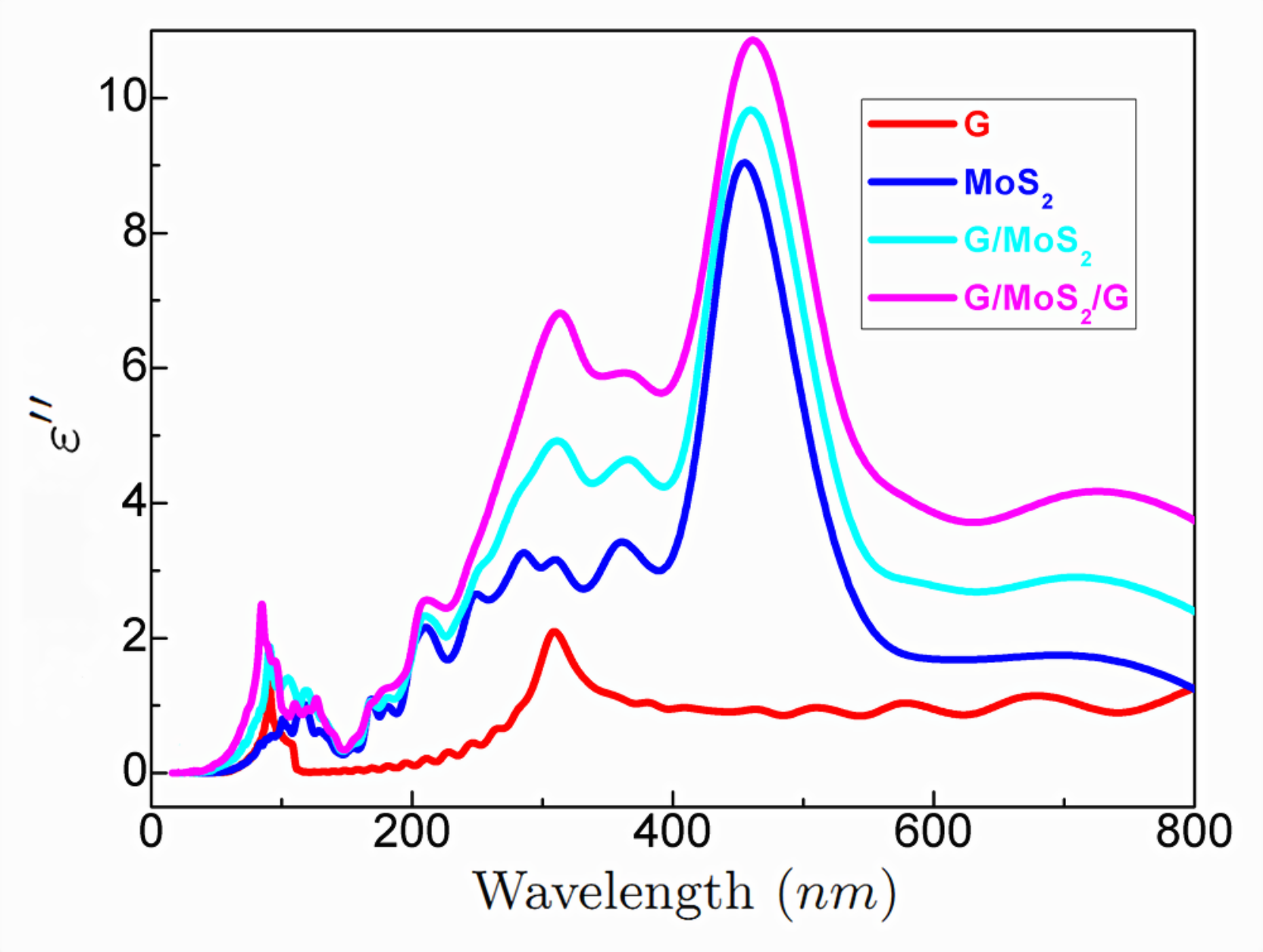}
\end{center}
\caption{(Color online) Imaginary part of dielectric function
($\varepsilon^{\prime\prime}$) of pristine graphene MoS$_2$
monolayers as well as corresponding hybrid G/MoS$_2$ and G/MoS$_2$/G
nanocomposites.}
\end{figure}

\section{Summary and Conclusions}

In summary, structural, electronic, electrical and optical
properties in hybrid G/MoS$_2$ and G/MoS$_2$/G nanocomposites are
studied via first-principles calculations. Graphene interacts
overall weakly with  via van der Waals interactions with their
intrinsic electronic properties preserved. Applying vertical
electric fields is very easy to induce tunable p-type doping of
graphene in hybrid G/MoS$_2$ nanocomposites and generate p-type and
n-type doping of graphene in hybrid G/MoS$_2$/G sandwiched
nanocomposites. Moreover, hybrid G/MoS$_2$ and G/MoS$_2$/G
nanocomposites display enhanced optical absorption compared to
simplex graphene and MoS$_2$ monolayers. With excellent electronic,
electrical and optical properties combined, ultrathin hybrid
graphene and MoS$_2$ nanocomposites are expected to be with great
applications in efficient electronic, electrochemical, photovoltaic,
photoresponsive and memory devices. \\

\section{ACKNOWLEDGMENTS}

This work is partially supported by the National Key Basic Research
Program (2011CB921404), by NSFC (21121003, 91021004, 2123307,
21222304), by CAS(XDB01020300). This work is also partially
supported by the Scientific Discovery through Advanced Computing
(SciDAC) program funded by U.S. Department of Energy, Office of
Science, Advanced Scientific Computing Research and Basic Energy
Sciences (W. H.). We thank the National Energy Research Scientific
Computing (NERSC) center, and the USTCSCC, SC-CAS, Tianjin, and
Shanghai Supercomputer Centers for the computational resources.

\end{document}